\documentclass[jsco]{academic}
\usepackage{amsmath}
\usepackage{amsfonts}
\usepackage{amssymb}
\usepackage{natbib}
\usepackage{fancyvrb}     
\newenvironment{mdisplay}
               {\fontsize{10}{12} \selectfont}

\usepackage{mapleenv} % still needed for mapletypeout environment

\MaplePromptString = {\raise 1pt \hbox{$\scriptstyle>$}}              % prompt
\MaplePromptSecondary = {\raise 1pt \hbox{$\phantom{\scriptstyle>}$}} % no prompts
\RecustomVerbatimEnvironment%
  {Verbatim}{Verbatim}
  {numbers=left,numbersep=5mm,frame=lines,framerule=0.4pt,%
   firstnumber=last,numberblanklines=false,fontfamily=tt,%
   fontsize=\footnotesize,commandchars=\\\{\},framesep=2pt}
\renewcommand{\FancyVerbFormatLine}[1]{\makebox[0cm][l]{${\hskip%
          -3.5ex}>$\kern -0.1ex #1}}
\def\today{December 10, 2002}
\def\nn{\nonumber \\}
\def\Id{\text{I\!d}}

\def\openZ{\mathbb{Z}}

\def\!{\kern -0.15ex}
\def\nn{\nonumber\\}
\def\CLIFFORD{\mbox{\bf \tt CLIFFORD}}
\def\BIGEBRA{\mbox{\bf \tt BIGEBRA}}

\def\Cli5plus{\mbox{\bf \tt Cli5plus}}

\def\cmulK{\mbox{\bf \tt cmul[K]}}

\def\LC{\mbox{\bf \tt LC}}

\def\linalg{\mbox{\bf \tt linalg}}

\def\mapop{\mbox{\bf \tt mapop}}
\def\mapoptwo{\mbox{\bf \tt mapop2}}
\def\map{\mbox{\bf \tt \&map}}
\def\contract{\mbox{\bf \tt contract}}
\def\vectorpart{\mbox{\bf \tt vectorpart}}
\def\Rop{\mbox{\bf \tt Rop}}
\def\EV{\mbox{\bf \tt EV}}
\def\gantipode{\mbox{\bf \tt gantipode}}

\newcommand{\ed}{\end{document}}

\begin{document}
\shortauthor{R. Ab{\l}amowicz and B. Fauser}
\shorttitle{Hopf algebras via $\BIGEBRA$ for Maple}
\title{Clifford and Gra\ss mann Hopf algebras via the $\BIGEBRA$ package 
for Maple$^{\text{(R)}}$%
\footnote{Package homepage is located at {\tt http://math.tntech.edu/rafal/} while 
$\CLIFFORD$ and $\BIGEBRA$ are available for Maple V, Maple 6, 7, 
and 8. Maple$^{\text{(R)}}$  is available from {\tt http://www.maplesoft.com}.}}
\author[1]{Rafa{\l} Ab{\l}amowicz}
\author[2]{Bertfried Fauser}
\address[1]{%
Department of Mathematics, Box 5054, Tennessee Technological University\\
Cookeville, TN 38505, USA, E-Mail: {\tt rablamowicz@tntech.edu}}
\address[2]{%
Universit\"at Konstanz, Fachbereich Physik, Fach M678, 78457 Konstanz\\
Germany, E-Mail: {\tt Bertfried.Fauser@uni-konstanz.de}}
\maketitle
\begin{abstract}
Hopf algebraic structures will replace groups and group representations as 
the leading paradigm in forthcoming times. $K$-theory, co-homology, 
entanglement, statistics, representation categories, quantized or twisted 
structures as well as more geometric topics of invariant theory, e.g., the 
Gra\ss mann-Cayley bracket algebra are all covered by the Hopf algebraic 
framework. The new branch of {\em experimental mathematics} allows one
to easily enter these fields through direct calculations using symbolic
manipulation and computer algebra system (CAS). We discuss problems which were
solved when building the $\BIGEBRA$ package for Maple and $\CLIFFORD$\footnote{The 
$\CLIFFORD$ package is described in a companion paper by
\citet{ablamowicz:fauser:2002c}} to handle tensor products, Gra\ss mann and 
Clifford algebras, coalgebras and Hopf algebras. Recent results showing 
the usefulness of CAS for investigating new and involved mathematics provide
us with examples. An outlook on further developments is given.
\end{abstract}

\section{Aim of the paper}

The first aim of this paper is to present the features of $\BIGEBRA,$ a Maple V rel.~5 
package. $\BIGEBRA$ was built to deal with tensored Clifford and Gra\ss mann algebras, and
it relies on $\CLIFFORD$ \citep{BIGEBRA,CLIFFORD}. While the emphasis here is
more on the package, we nevertheless provide novel results in the last section
on quantum Yang-Baxter equations derived from a Clifford bi-convolution.
$\CLIFFORD$ and $\BIGEBRA$ are meanwhile available for Maple 6, 7 and Maple 8.

Secondly, we will not pass an opportunity to raise the flag for the subject
of {\em experimental mathematics}. We strongly believe that experimental mathematics
based on algorithmic approach has already been been changing research and teaching of mathematics and theoretical physics 
\citep{fauser:2002c}. The main aspects of experimental mathematics, in our view, are as follows:

\vskip 5pt
\noindent
{\bf Computations}
\vskip 5pt

Various results have been achieved by brute force calculations. CAS simplifies difficult 
computations with non-commutative algebras, and it allows one to tackle problems 
impossible for hand calculations. It makes fewer errors and its results can be validated.  
A CAS {\em cannot} and {\em does not} replace knowledge of the mathematics behind the problem 
and it requires that all concepts be well defined and sound.

\vskip 5pt
\noindent
{\bf Checking Assertions}
\vskip 5pt

Having a CAS at hand, one can easily check one's own assertions. A single counter-example uncovers one's misconceptions and leads to a more sound understanding of the subject. Indeed, when checking a {\em theorem} on examples, one can occasionally find that it is not valid or that it has not been correctly formulated. This in turn helps with finding errors in the computer code ({\em Never trust any result generated by a computer!}) and/or with coming up with a proper formulation of such ill-stated theorem.

\vskip 5pt
\noindent
{\bf Developing New Mathematics}
\vskip 5pt

It has proved to be necessary to develop new mathematics for solving problems in physics. For example, we have introduced the {\em Hopf Gebra} as a weakened form of a {\em Hopf Algebra}. Using a CAS this was done by first exploring the set of principles which yielded correct physical results and then by fixing the mathematics of the new structure.

\vskip 5pt
\noindent
{\bf Teaching}
\vskip 5pt

Having a well developed CAS, it can be useful in teaching! Students can check their own 
prejudices by exemplifying them directly on a computer. This requires a stable code which 
does not allow for illegal input, etc. $\CLIFFORD'$s code is stable while $\BIGEBRA$ still 
needs knowledge of the user because to gain speed, type and input checking is hardly done 
since it is time consuming. Hence, $\CLIFFORD$ (and with restrictions $\BIGEBRA)$ can be 
effectively used for a fast access to the mathematical field of Gra\ss mann and Clifford 
algebras and cogebras in teaching. Students can make their own experiments and develop --given a 
CAS which is stable under silly input-- a sound understanding of the mathematical structures 
in question.  Teaching mathematics or physics using a CAS allows one to abstract from technical 
details on the {\em first} approach. However, if the goal is that students develop an 
understanding of the topic, they should be encouraged to recalculate by hand and to recode 
themselves. Otherwise no deep understanding will ensue. A good example how this can be 
achieved is \citet{wright:2002a}. 

\vskip 5pt
\noindent
{\bf Experimental Mathematics}
\vskip 5pt

Having the opportunity to deal with a CAS opens the field of {\em experimental mathematics}. This includes partly the other
topics given above but it should not be underestimated due to its own dynamics. Exploring mathematics {\em by doing particular
experiments} and by strengthening or weakenning once's \textit{own} assumptions is of extreme value as it allows one to enter a new mathematical field quickly and in a secure way.

\vskip 5pt
\noindent
{\bf Algorithmic Understanding}
\vskip 5pt

We share a growing belief \citep{gruel:pfister:2002} that computer usage not only allows one to successfully complete non-trivial computations but also, through the developments of algorithms, contributes to a better understanding of the problem. The so 
called \textit{algorithmic approach} leads to achieving a computational proficiency that in turn leads to a deeper mastery of the subject on the theoretical level. 

\section{The $\BIGEBRA$ package}

We concentrate in this article on $\BIGEBRA$ and its features which generally go beyond those of $\CLIFFORD$ and Maple. We assume that the reader is at somewhat familiar with the syntax and features of Maple, see, e.g., \citet{wright:2002a}, while $\CLIFFORD$ is described in \citep{ablamowicz:fauser:2002c}.

The package loads in a non verbose (silent) mode with the following commands while the linear algebra package $\linalg$ is loaded for convenience only. $\BIGEBRA$ automatically loads $\CLIFFORD$ which is needed for its functionality:
\begin{Verbatim}
restart:_CLIENV[_SILENT]:=true:with(linalg):with(Bigebra);
\end{Verbatim}
\vskip-12pt
\begin{mdisplay}
\begin{setlength}{\multlinegap}{0pt}
\begin{multline*}
[\mathit{\&cco}, \,\mathit{\&gco}, \,\mathit{\&gco\_d},\,\mathit{\&gpl\_co}, \,\mathit{\&map}, \,\mathit{\&v}, \,
\mathit{EV}, \,\mathit{VERSION}, \,\mathit{bracket}, \,\mathit{cco\_monom}, \,\mathit{contract}, \\
\mathit{do\_compose}, \,\mathit{drop\_t}, \,\mathit{eps}, \,\mathit{gantipode}, \,\mathit{gco\_d\_monom}, \,
\mathit{gco\_monom}, \,\mathit{gco\_unit}, \,\mathit{gpl\_co\_monom},  \\
\mathit{gpl\_co\_monom2}, \,\mathit{gswitch}, \,\mathit{init}, \,\mathit{linop}, \,\mathit{linop2}, \,\mathit{lists2mat}, \,
\mathit{lists2mat2}, \,\mathit{make\_BI\_Id}, \,\mathit{mapop},\\
\mathit{mapop2}, \,\mathit{meet}, \,\mathit{op2mat}, \,\mathit{op2mat2}, \,\mathit{pairing}, \,\mathit{peek},\,
\mathit{poke}, \,\mathit{remove\_eq}, \,\mathit{switch}, \,\mathit{tcollect}, \,\mathit{tsolve1},  \\
\mathit{type/tensorbasmonom}, \,\mathit{type/tensormonom},\,\mathit{type/tensorpolynom}]
\end{multline*}
\end{setlength}
\end{mdisplay}
\vskip-12pt
\noindent
The output is a list with available functions; some of them are for internal use only. $\CLIFFORD$ and $\BIGEBRA$ come with an extensive 
online help page system which is included in the Maple online help and can be searched. It contains not only the syntax of the 
procedures but even a good deal of unpublished mathematics. It is meant to present to the user the mathematical concepts at hand. Help
topics may be reached in Maple fashion with the command \verb+?Bigebra+ followed by the enter key.

\subsection{Tensor product}

Given the functionality of $\CLIFFORD$ to compute with Clifford algebras, we need two more key features to enter the realm of bi- and Hopf algebras. The first one is the tensor product and some functions to manipulate it with while the second is the coproduct. The latter, however, follows naturally from the algebra structure on the linear dual space \citep{milnor:moore:1965a}.

Maple comes with a \verb+define+ facility which allows to introduce ampersand operators \verb+&<name>+ that are associative (flat), commutative (orderless), linear or multilinear. This facility unfortunately has two drawbacks:
\begin{itemize}
\item 
It cannot deal with user-defined scalars (ring elements). This is mathematically insensitive since any definition of a linear or multilinear
function {\em needs} linearity with respect to ``scalars'' that one wants to compute with. We want to consider, e.g., tensor products (multilinear associative) over arbitrary rings, e.g., the integers, polynomial rings etc.
\item
It produces code which gives wrong output. 
\end{itemize}
Remembering that Gra\ss mann basis multivectors are denoted as \verb+Id+, \verb+e1+, \verb+e2+, \verb+ei+, \verb+e1we2+, \verb+e1we3+,
\verb+eiwej,...+, let us explore properties of the associative (flat) multilinear operator \verb+&r+ (comments in the code are separated by 
\verb+#+):
\begin{Verbatim}
restart:                          # unload BIGEBRA !
define(`&r`,flat,multilinear):    # produce the operator `&r`
out[1]:=&r(2*e1,2*e2+4*a*e3):     # a is not known to be a ring element
out[2]:=e1+e2 &r e3, &r(e1+e2,e3):# infix form is dangerous
out[1];out[2];                    # we delete output for typographical reasons
\end{Verbatim}
\begin{mdisplay}
\[
4\,(\mathit{e1}\,\mathrm{\&r}\,\mathit{e2}) + 8\,(\mathit{e1}\,
\mathrm{\&r}\,a\,\mathit{e3})
\]
\[
\mathit{e1} + (\mathit{e2}\,\mathrm{\&r}\,\mathit{e3}), \,(
\mathit{e1}\,\mathrm{\&r}\,\mathit{e3}) + (\mathit{e2}\,\mathrm{
\&r}\,\mathit{e3})
\]
\end{mdisplay}
Notice that Maple does not know how to deal with $a$ so, effectively, the multilinearity is established over the integers only. If one wants to compute over a polynomial ring or, more generally, with $K$-modules, this function is useless.
\begin{Verbatim}
constants:=constants,a:
out[1]:=&r(2*e2,2*e1+4*a*e1): # a is known to be a constant now
out[2]:=&r(true*e1,false*e2): # true and false are 'constants'
constants;out[1];out[2];
\end{Verbatim}
\begin{mdisplay}
\[
\mathit{false}, \,\gamma , \,\infty , \,\mathit{true}, \,\mathit{
Catalan}, \,\mathit{FAIL}, \,\pi , \,a
\]
\[
4\,(\mathit{e2}\,\mathrm{\&r}\,\mathit{e1}) + 8\,a\,(\mathit{e2}
\,\mathrm{\&r}\,\mathit{e1})
\]
\[
\mathit{true}\,\mathit{false}\,(\mathit{e1}\,\mathrm{\&r}\,
\mathit{e2})
\]
\end{mdisplay}
The last line does not make sense at all. Furthermore we find these peculiarities: 
\begin{Verbatim}
out[1]:=&r(-e1,-e2):         # -> &r(e1,e2)
out[2]:=&r(-e1):             # -> -1*&r(e1)
out[3]:=&r(2.5*e1):          # -> 2.5*&r(e1)
out[4]:=&r(&r(-e1)):         # -> &r(e1) 
out[1],out[2],out[3],out[4];
\end{Verbatim}
\begin{mdisplay}
\[
( - \mathit{e1})\,\mathrm{\&r}\,( - \mathit{e2}), \, - (1\,
\mathrm{\&r}\,\mathit{e1}), \,2.5\,(1\,\mathrm{\&r}\,\mathit{e1})
, \, - \mathrm{\&r}(1, \,1, \,\mathit{e1})
\]
\end{mdisplay}
While the first expression is not even simplified, the last three results are just plain wrong. This error was detected in Maple 5 and is partially still present up to Maple 7. Hence, $\BIGEBRA$ includes a patched \verb+define+ facility.

In $\CLIFFORD$ and $\BIGEBRA$ almost everything can be treated as scalars, hence one can compute over function spaces etc. The generic tensor product of $\BIGEBRA$ is given by \verb+&t+.
\begin{Verbatim}
# reload BIGEBRA, defined will be patched
restart:_CLIENV[_SILENT]:=true:with(linalg):with(Bigebra):  
out[1]:=e1+e2 &t e3, &t(e1+e2,e3):   # infix form is still dangerous!
out[2]:=&t(&t(-e1)):                 # fixed
out[3]:=&t(2.5*e1):
out[4]:=&t(a*e1+2*e2,sin(phi)*e3):
out[1];out[2],out[3];out[4];
\end{Verbatim}
\begin{mdisplay}
\[
\mathit{e1} + (\mathit{e2}\,\mathrm{\&t}\,\mathit{e3}), \,(
\mathit{e1}\,\mathrm{\&t}\,\mathit{e3}) + (\mathit{e2}\,\mathrm{
\&t}\,\mathit{e3})
\]
\[
 - \mathrm{\&t}(\mathit{e1}), \,2.5\,\mathrm{\&t}(\mathit{e1})
\]
\[
a\,\mathrm{sin}(\phi )\,(\mathit{e1}\,\mathrm{\&t}\,\mathit{e3})
 + 2\,\mathrm{sin}(\phi )\,(\mathit{e2}\,\mathrm{\&t}\,\mathit{e3
})
\]
\end{mdisplay}
The patched \verb+define+ which ships with $\BIGEBRA$ can handle tensor products over quite general rings. We give a final example where we define a tensor product \verb+&r+ over the polynomial ring $\openZ[[x]]$ with integer coefficients.
\begin{Verbatim}
define(`&r`,flat,multilinear):
unprotect(`type/cliscalar`):             # we want to change this type
`type/cliscalar`:=proc(expr) type(expr,polynom(integer,x)) end:
out[1]:=type( 2*x+3*x^5+5,cliscalar):    # true, since in Z[[x]]
out[2]:=type( x/2        ,cliscalar):    # false, fraction
out[3]:=type( 2*y^2+3*y-1,cliscalar):    # false, wrong indeterminant
out[1],out[2],out[3];
\end{Verbatim}
\begin{mdisplay}
\[
\mathit{true}, \,\mathit{false}, \,\mathit{false}
\]
\end{mdisplay}
Now we are ready to use the new tensor product \verb+&r+ over $\openZ[[x]]:$
\begin{Verbatim}
out[1]:=&r(2.5*x,4*y):
out[2]:=&r(3*x^2-x+5,x-x^4):
out[3]:=&r(3*y*x^2-x-5,x*y-x^4):
out[1];out[2];out[3];
\end{Verbatim}
\begin{mdisplay}
\[
4\,x\,(2.5\,\mathrm{\&r}\,y)
\]
\[
3\,x^{3}\,(1\,\mathrm{\&r}\,1) - 3\,x^{6}\,(1\,\mathrm{\&r}\,1)
 - x^{2}\,(1\,\mathrm{\&r}\,1) + x^{5}\,(1\,\mathrm{\&r}\,1) + 5
\,x\,(1\,\mathrm{\&r}\,1) - 5\,x^{4}\,(1\,\mathrm{\&r}\,1)
\]
\[
3\,x^{3}\,(y\,\mathrm{\&r}\,y) - 3\,x^{6}\,(y\,\mathrm{\&r}\,1)
 - x^{2}\,(1\,\mathrm{\&r}\,y) + x^{5}\,(1\,\mathrm{\&r}\,1) - 5
\,x\,(1\,\mathrm{\&r}\,y) + 5\,x^{4}\,(1\,\mathrm{\&r}\,1)
\]
\end{mdisplay}
Observe, that real numbers like $2.5$ or the variable $y$ are still not treated as scalars but integers and $x$ are. 

\subsection{Basic tools}

Having defined a tensor product, we need to have tools for operating on such structures. Let us fix notation as follows: A single term having no prefactors \verb+&t(b1,...,bn)+, where $b_1,\ldots,b_n$ are Gra\ss mann basis multivectors, will be called a {\em tensor basis monom} or a {\em word} that is composed from letters of the Gra\ss mann multivector alphabet. {\em A tensor monom} is a tensor basis monom with a ``scalar'' prefactor where ``scalar'' means a ring element, a function, a polynomial, or just a number. {\em A tensor polynom}, or just a tensor, is a sum of tensor monoms. It is the multilinearity that guarantees that every tensor can be written as a linear combination of some tensor basis monoms (Hilbert basis theorem). The $i$-th place in a list of arguments of a tensor will be called the $i$-th {\em slot}. 

Acting on a tensor means:
\begin{itemize}
\item {\em Removing or grafting of terms from or into a tensor}
\item {\em Rearranging the tensor}
\item {\em Acting with operators on $n$-slots producing $m$-slots}
\end{itemize}
The first set of operations is given by \verb+peek+ and \verb+poke+. They work as expected: \verb+peek+ takes as an argument a number of the slot which it intends to remove and returns a sequence of lists of the removed terms and the remaining tensor. Procedure \verb+poke+ needs three arguments: a tensor, a Gra\ss mann multivector which will be put into the $i$-th place, and the slot number $i.$
\begin{Verbatim}
f:=x-> 'peek'(x,2):x:=&t(e1,a*e2+b*e3,e3):f(x)=eval(f(x));
\end{Verbatim}
\begin{mdisplay}
\[
\mathrm{peek}(a\,\mathrm{\&t}(\mathit{e1}, \,\mathit{e2}, \,
\mathit{e3}) + b\,\mathrm{\&t}(\mathit{e1}, \,\mathit{e3}, \,
\mathit{e3}), \,2)=([a\,\mathit{e2}, \,\mathit{e1}\,\mathrm{\&t}
\,\mathit{e3}], \,[b\,\mathit{e3}, \,\mathit{e1}\,\mathrm{\&t}\,
\mathit{e3}])
\]
\end{mdisplay}
\begin{Verbatim}
g:=x-> 'poke'(x,e2,3):x:=&t(e1,a*e2+b*e3):g(x)=eval(g(x));
\end{Verbatim}
\begin{mdisplay}
\[
\mathrm{poke}(a\,(\mathit{e1}\,\mathrm{\&t}\,\mathit{e2}) + b\,(
\mathit{e1}\,\mathrm{\&t}\,\mathit{e3}), \,\mathit{e2}, \,3)=a\,
\mathrm{\&t}(\mathit{e1}, \,\mathit{e2}, \,\mathit{e2}) + b\,
\mathrm{\&t}(\mathit{e1}, \,\mathit{e3}, \,\mathit{e2})
\]
\end{mdisplay}
The second group of operations consists of switches --also called crossings or braids-- which allow to reorder the tensors. From a CAS point, it would be easy, say, to multiply the $i$-th and the $j$-th slots of a tensor and place the output into the $k$-th slot. However, mathematical reasons disallow such a brute method. If one demands that any reordering process can be generated from the reordering rules of generators, one has to assume that a crossing is a {\em natural transformation} in a functorial sense and obeys coherence, 
see \citet{kelly:laplaza:1980a,lyubashenko:1995a, lyubashenko:1995b}. In fact these two conditions imply the quantum Yang Baxter equation which is discussed below. The Gra\ss mann Hopf algebraic graded crossing and the ordinary switch (swap) of two elements fulfill these properties.

The \verb+switch+ procedure swaps two {\em adjacent} tensor slots in a general tensor polynomial, while the graded switch \verb+gswitch+ respects the grading of the factors. Both functions take as a second argument $i,$ the number of the tensor slot to act on, that is,  to switch the $i$-th and the $(i+1)$-st tensor entries.
\begin{Verbatim}
f1:=x->'switch'(x,2):x:=&t(e1,e2,e3,e4):f1(x)=eval(f1(x));
\end{Verbatim}
\begin{mdisplay}
\[
\mathrm{switch}(\mathrm{\&t}(\mathit{e1}, \,\mathit{e2}, \,
\mathit{e3}, \,\mathit{e4}), \,2)=\mathrm{\&t}(\mathit{e1}, \,
\mathit{e3}, \,\mathit{e2}, \,\mathit{e4})
\]
\end{mdisplay}
\begin{Verbatim}
f2:=x->'gswitch'(x,2):x:=&t(e1,e2,e3,e4):f2(x)=eval(f2(x));
\end{Verbatim}
\begin{mdisplay}
\[
\mathrm{gswitch}(\mathrm{\&t}(\mathit{e1}, \,\mathit{e2}, \,
\mathit{e3}, \,\mathit{e4}), \,2)= - \mathrm{\&t}(\mathit{e1}, \,
\mathit{e3}, \,\mathit{e2}, \,\mathit{e4})
\]
\end{mdisplay}
\begin{Verbatim}
x:=&t(e1,e2,e3we4,e5):f2(x)=eval(f2(x));
\end{Verbatim}
\begin{mdisplay}
\[
\mathrm{gswitch}(\mathrm{\&t}(\mathit{e1}, \,\mathit{e2}, \,
\mathit{e3we4}, \,\mathit{e5}), \,2)=\mathrm{\&t}(\mathit{e1}, \,
\mathit{e3we4}, \,\mathit{e2}, \,\mathit{e5})
\]
\end{mdisplay}
Notice the different signs for \verb+switch+ and \verb+gswitch+ depending in the latter morphism on the grading of the Gra\ss mann multivector elements.

Furthermore, $\BIGEBRA$ allows to act with any operator on tensors. We distinguish such operations by the number of input and output slots. An ordinary endomorphism is a $1\rightarrow 1$ map while, e.g., a product is a $2\rightarrow 1$ map. Functions which are currently available are $\mapop$ and $\mapoptwo$ for $1\rightarrow 1$ and $2\rightarrow 2$ operators, $\map$ for $2\rightarrow 1$ operators and $\contract$ for $2\rightarrow 0$ operators. Let us first show a linear operator $proj1$ acting on a certain tensor slot:
\begin{Verbatim}
proj1:=proc(x) vectorpart(x,1) end:
L:=x->'mapop'(x,2,proj1):x:=&t(Id,e1+e1we2,e3):L(x)=eval(L(x));
\end{Verbatim}
\begin{mdisplay}
\[
\mathrm{mapop}(\mathrm{\&t}(\mathit{Id}, \,\mathit{e1}, \,
\mathit{e3}) + \mathrm{\&t}(\mathit{Id}, \,\mathit{e1we2}, \,
\mathit{e3}), \,2, \,\mathit{proj1})=\mathrm{\&t}(\mathit{Id}, \,
\mathit{e1}, \,\mathit{e3})
\]
\end{mdisplay}
where $\vectorpart(x,n)$ projects on the $n$-vector part of $x$ in the Gra\ss mann basis. Let us define a general element $x$ in a Gra{\ss}mann or Clifford algebra over a two dimensional vector space and a general operator $\Rop$ given by its matrix $R$ in a co-contravariant canonical basis:
\begin{Verbatim}
restart:_CLIENV[_SILENT]:=true:with(linalg):with(Bigebra):
dim_V:=2:                                 # set dimension to 2
bas:=cbasis(dim_V):                       # generate a list of basis monoms
X:=add(x[i]*bas[i],i=1..2^dim_V):         # general element
R:=matrix(2^dim_V,2^dim_V,(i,j)->r[i,j]): # matrix with entries r[i,j]
Rop:=proc(x) linop(x,R) end:              # the operator Rop
X;evalm(R);
\end{Verbatim}
\begin{mdisplay}
\[
{x_{1}}\,\mathit{Id} + {x_{2}}\,\mathit{e1} + {x_{3}}\,\mathit{e2
} + {x_{4}}\,\mathit{e1we2}
\]
\[
 \left[ 
{\begin{array}{cccc}
{r_{1, \,1}} & {r_{1, \,2}} & {r_{1, \,3}} & {r_{1, \,4}} \\
{r_{2, \,1}} & {r_{2, \,2}} & {r_{2, \,3}} & {r_{2, \,4}} \\
{r_{3, \,1}} & {r_{3, \,2}} & {r_{3, \,3}} & {r_{3, \,4}} \\
{r_{4, \,1}} & {r_{4, \,2}} & {r_{4, \,3}} & {r_{4, \,4}}
\end{array}}
 \right] 
\]
\end{mdisplay}
The action of $\Rop$ on Gra\ss mann multivectors and tensors is computed as
\begin{Verbatim}
out[1]:=Rop(Id):out[2]:=Rop(X):out[3]:=mapop(&t(e1,Id,e2),2,Rop):
out[1];out[2];out[3];
\end{Verbatim}
\vskip-10pt
\begin{mdisplay}
\begin{gather*}
{r_{1, \,1}}\,\mathit{Id} + {r_{2, \,1}}\,\mathit{e1} + {r_{3, \,
1}}\,\mathit{e2} + {r_{4, \,1}}\,\mathit{e1we2},
\end{gather*}
\vskip-33pt
\begin{setlength}{\multlinegap}{0pt}
\begin{multline*}
({r_{1, \,1}}\,{x_{1}} + {r_{1, \,2}}\,{x_{2}} + {r_{1,\,3}}\,{x_{3}} + {r_{1,\,4}}\,{x_{4}})\,\mathit{Id} + 
({r_{3, \,4}}\,{x_{4}} + {r_{3, \,1}}\,{x_{1}} + {r_{3, \,2}}\,{x_{2}} + {r_{3, \,3}}\,{x_{3}})\,\mathit{e2} \\
+ ({r_{2, \,2}}\,{x_{2}} + {r_{2, \,3}}\,{x_{3}} + {r_{2, \,4}}\,{x_{4}} + {r_{2, \,1}}\,{x_{1}})\,\mathit{e1} + 
({r_{4, \,1}}\,{x_{1}} + {r_{4, \,2}}\,{x_{2}} + {r_{4, \,3}}\,{x_{3}} + {r_{4, \,4}}\,{x_{4}})\,\mathit{e1we2},
\end{multline*}
\end{setlength}
\vskip-37pt
\begin{gather*}
{r_{1, \,1}}\,\mathrm{\&t}(\mathit{e1}, \,\mathit{Id}, \,\mathit{e2}) + {r_{2, \,1}}\,\mathrm{\&t}(\mathit{e1}, \,\mathit{e1}, \,
\mathit{e2}) + {r_{3, \,1}}\,\mathrm{\&t}(\mathit{e1}, \,\mathit{e2}, \,\mathit{e2}) + {r_{4, \,1}}\,\mathrm{\&t}(\mathit{e1}, \,
\mathit{e1we2}, \,\mathit{e2})
\end{gather*}
\end{mdisplay}
\vskip-5pt
Operators may be defined as any Maple procedures also. 

The next class of operations is the application of product maps, i.e., $2\rightarrow 1,$ which have two {\em adjacent} entry slots and one output slot. Examples of such operations which are predefined in $\BIGEBRA$ are the Gra\ss mann and the Clifford products, and the contractions. User defined functions may be applied also. They are mapped onto tensors via the $\map$ function which takes a tensor, the slot number and the (product) function as input parameters:
\begin{Verbatim}
trm:=&t(Id,e1,e2,Id):
out[1]:=`´&map(&t(Id,e1,e2,Id),2,wedge)´`,` --> `,&map(trm,2,wedge):
out[2]:=`´&map(&t(Id,e1,e2,Id),2,cmul[K])´`,` --> `,&map(trm,2,cmul[K]):
out[3]:=`´&map(&t(Id,e1,e2,Id),2,LC,K)´`,` --> `, &map(trm,2,LC,K):
out[1];out[2];out[3];
\end{Verbatim}
\begin{mdisplay}
\[
\mathit{\char"B4\&map(\&t(Id,e1,e2,Id),2,wedge)\char"B4}, \,\mathit{ --> }, 
\,\mathrm{\&t}(\mathit{Id}, \,\mathit{e1we2}, \,\mathit{Id})
\]
\[
\mathit{\char"B4\&map(\&t(Id,e1,e2,Id),2,cmul[K])\char"B4}, \,\mathit{ --> }, \,
\mathrm{\&t}(\mathit{Id}, \,\mathit{e1we2}, \,\mathit{Id}) + {K_{
1, \,2}}\,\mathrm{\&t}(\mathit{Id}, \,\mathit{Id}, \,\mathit{Id})
\]
\[
\mathit{\char"B4\&map(\&t(Id,e1,e2,Id),2,LC,K)\char"B4}, \,\mathit{ --> }, \,
{K_{1, \,2}}\,\mathrm{\&t}(\mathit{Id}, \,\mathit{Id}, \,\mathit{Id})
\]
\end{mdisplay}
Note that the Clifford product $\cmulK$ and the left contraction $\LC(...,K)$ depend on an arbitrary bilinear form $K$ which is then passed on as an additional parameter to the procedure $\map.$ This mechanism can then be used to deal with different Clifford algebras in different tensor slots.

As a last example of actions on tensors we examine the evaluation and other $2 \rightarrow 0$ mappings with values in the base ring. A very important case is given by the {\em evaluation} which employs the action of dual elements w.r.t. the canonical basis on  multivectors. In fact one would need a new kind of basis vectors, however, for technical reasons of the current version of $\BIGEBRA,$ it is the {\em user} who has to take responsibility for keeping track of the tensor slots which are to contain co(multi)vectors.
\begin{Verbatim}
out[1]:=[EV(Id,Id),EV(e1,e2),EV(e1,e1)]: # eval = Kronecker delta
out[2]:=`´contract´(&t(Id,e1,e1+e2+e3,Id),2,EV)`,
        `  -->  `,contract(&t(Id,e1,e1+e2+e3,Id),2,EV):
out[1];out[2];
\end{Verbatim}
\begin{mdisplay}
\[
[1, \,0, \,1]
\]
\[
\mathit{\char"B4contract\char"B4(\&t(Id,e1,e1+e2+e3,Id),2,EV)}, \,\mathit{\ \ 
-->\ \ }, \,\mathit{Id}\,\mathrm{\&t}\,\mathit{Id}
\]
\end{mdisplay}

\subsection{Gra\ss mann Hopf algebra}

A Gra\ss mann Hopf algebra is a vector space that is a Gra\ss mann algebra and a Gra\ss mann coalgebra fulfilling certain compatibility laws \citep{sweedler:1969a}. Since the algebra side is well known, we give some explanation on the coalgebra side. A coproduct on an algebra can be defined as the categorial dual of a product via a linear form defined on the vector space. This law is called {\em product coproduct duality} \citep{milnor:moore:1965a}. Now, it is easy to check that covectors may form a covector Gra\ss mann algebra $V^{\vee},$ where $\vee$ is the Gra\ss mann exterior product on co-multivectors. Using duality via the evaluation map $\langle . \mid . \rangle$  we find
\begin{align}
\langle w \mid x\wedge y\rangle &= 
\langle w_{(1)}\mid y\rangle
\langle w_{(2)}\mid x\rangle, \nn
\langle w \vee w^\prime \mid x\rangle &= 
\langle w\mid x_{(2)}\rangle
\langle w^\prime \mid x_{(1)}\rangle
\end{align}
where we have used Sweedler's notation for the coproduct $\Delta(x) = \sum_{(x)} x_{(1)}\otimes x_{(2)}$ (the summation symbol is usually dropped). The evaluation map is given by the function $\EV$ from the package. The coproduct is seen to be a split of the homogeneous
multivectors into pairs where the sign of permutation is taken into account. We give one example with symbolic indices $a$ and $b:$
\begin{Verbatim}
&gco(eaweb);
\end{Verbatim}
\begin{mdisplay}
\[
(\mathit{Id}\,\mathrm{\&t}\,\mathit{eaweb}) + (\mathit{ea}\,
\mathrm{\&t}\,\mathit{eb}) - (\mathit{eb}\,\mathrm{\&t}\,\mathit{
ea}) + (\mathit{eaweb}\,\mathrm{\&t}\,\mathit{Id})
\]
\end{mdisplay}
The first compatibility law valid in a Hopf algebra is that the product is an coalgebra homomorphism and the coproduct is an algebra homomorphism, see \citet{fauser:2002c}. Furthermore, an antipode exists, which in the Gra\ss mann case turns out to be isomorphic to the 
grade involution. We check this in dimension~2:
\begin{Verbatim}
dim_V:=2:
`S^`=matrix(2^dim_V,2^dim_V,(i,j)->EV(bas[i],gantipode(bas[j])));
`Grade_Inv`=matrix(2^dim_V,2^dim_V,(i,j)->EV(bas[i],gradeinv(bas[j])));
\end{Verbatim}
\begin{mdisplay}
\begin{gather*}
\mathit{S\symbol{94}}= \left[ 
\begin{array}{rrrr}
1 & 0 & 0 & 0 \\
0 & -1 & 0 & 0 \\
0 & 0 & -1 & 0 \\
0 & 0 & 0 & 1
\end{array}
\right], \quad
\mathit{Grade\_Inv}= \left[ 
\begin{array}{rrrr}
1 & 0 & 0 & 0 \\
0 & -1 & 0 & 0 \\
0 & 0 & -1 & 0 \\
0 & 0 & 0 & 1
\end{array}
 \right] 
\end{gather*}
\end{mdisplay}
The above described tools allow us to perform all algebraic manipulations in a Gra\ss mann Hopf algebra, also using symbolic indices. 

\subsection{Clifford bi-convolution}
A {\em Clifford bi-convolution} is a space endowed with a Clifford algebra structure and a Clifford coalgebra structure. Both products are unital and associative but not all such convolutions posses an antipode. However if an antipode exists then one can prove that one deals then with a Clifford Hopf algebra \citep{oziewicz:2001b,fauser:2002c}. We are interested in Clifford algebras over a general bilinear form, hence we can introduce the Clifford product via the Chevalley deformation \citep{chevalley:1997a}. Let $x$ be an element in $V,$ the space of generators, and let $u$ be a general element in $V^\wedge (= \bigwedge V).$ Then, one defines the action of $x$ on $U$ as
\[
\gamma_{x}\,\circ\,u = i_{x}(u) + x \wedge u
\]
where $i_{x}$ is the {\em inner product} or {\em contraction}. The action is then extended to a map $\circ : V \otimes V \mapsto V$ by demanding that
\[
i_{x}(y) = B(x,y),\qquad i_{u\wedge v}(w) = i_{u}(i_{v}(w))
\]
\[
i_{x}(u\wedge v) = i_{x}(u)\wedge v + \hat{u} \wedge i_{x}(v)
\]
with $\hat{u}= (-1)^{\partial u}u$ being the grade involution, e.g., \citet{fauser:2002c}. It is a remarkable fact that this product can be directly defined via a deformation called {\em cliffordization} by Rota and Stein \citep{rota:stein:1994a}. It may also be called a Drinfeld twist, see below. Using the undeformed Gra\ss mann product $\wedge$ and coproduct $\Delta$ one can write 
\[
u\,\circ\,v = B^{\wedge}(u_{(2)},v_{(1)})\, u_{(1)}\wedge v_{(2)}
\]
Here $B^{\wedge}$ is the scalar valued bilinear form tied to the inner product action via the counit $\epsilon$ as $B^{\wedge}(a,b) = \epsilon(i_{a}(b)).$ We exemplify this in the $\BIGEBRA$ package as
\begin{Verbatim}
restart:_CLIENV[_SILENT]:=true:with(Bigebra):
unprotect(gamma):
Gamma:=proc(U) local x; x:=op(procname); LC(x,U)+wedge(x,U) end:
out[1]:=gamma[e1](Id)=Gamma[e1](Id):
out[2]:=gamma[e1](e2)=Gamma[e1](e2):
out[3]:=gamma[e1](e2we3)=Gamma[e1](e2we3):
out[1],out[2],out[3];
\end{Verbatim}
\begin{mdisplay}
\[
{\gamma _{\mathit{e1}}}(\mathit{Id})=\mathit{e1}, \,{\gamma _{
\mathit{e1}}}(\mathit{e2})={B_{1, \,2}}\,\mathit{Id} + \mathit{
e1we2}, \,{\gamma _{\mathit{e1}}}(\mathit{e2we3})={B_{1, \,2}}\,
\mathit{e3} - {B_{1, \,3}}\,\mathit{e2} + \mathit{e1we2we3}
\]
\end{mdisplay}
If we do the same using the cliffordization process, which we will make explicit by defining a function \verb+cliff+ as to avoid using the internal function \verb+cmul+, we get
\begin{Verbatim}
cliff:=proc(x,y) clicollect(simplify(drop_t(&map(contract(
                         &t(&gco(x),&gco(y)),2,scalarpart@LC),1,wedge)))) end:
out[1]:=e1    &C e2we3 = cliff(e1   ,e2we3):
out[2]:=e1we2 &C e2we3 = cliff(e1we2,e2we3):
out[1];out[2];
\end{Verbatim}
\begin{mdisplay}
\[
\mathit{e1}\,\mathrm{\&C}\,\mathit{e2we3}={B_{1, \,2}}\,\mathit{
e3} - {B_{1, \,3}}\,\mathit{e2} + \mathit{e1we2we3}
\]
\[
\mathit{e1we2}\,\mathrm{\&C}\,\mathit{e2we3}={B_{2, \,2}}\,
\mathit{e1we3} - {B_{1, \,2}}\,\mathit{e2we3} - {B_{2, \,3}}\,
\mathit{e1we2} - ( - {B_{2, \,2}}\,{B_{1, \,3}} + {B_{2, \,3}}\,{
B_{1, \,2}})\,\mathit{Id}
\]
\end{mdisplay}

The most interesting structure which we can now access computationally with the $\BIGEBRA$ package is that of a Clifford bi-convolution. In any pair ($U,V)$ of structures having a product on $V\otimes V$ and a coproduct on $U,$ which need not to be a bialgebra or Hopf algebra 
at all, one can define a convolution product between morphisms $f,g,... : U \rightarrow V,$ see \citet{fauser:2002c}. We consider here the Gra\ss mann Hopf algebra and endomorphisms, and define the {\em star} or {\em convolution product} of endomorphisms according to Sweedler,
\citep{sweedler:1969a} as
\[
(f \star g)(x) = m \circ (f\otimes g)\circ \Delta(x) 
\]
where $\circ$ is the composition of maps. The element $x$ can be dropped safely. Re-examining the definition of the antipode shows that the map is the convolutive inverse of the identity morphism $\Id_V$ on $V.$ In the following steps we will compute the antipodes of the
Gra\ss mann Hopf algebra and of a twisted Gra\ss mann Hopf algebra where only the product is deformed, and of the Clifford bi-convolution, where both structure maps, product and coproduct have been deformed. The last possibility is beyond current deformation theory, where 
only one structure map is deformed. We choose dimension 2 for the generating space which gives a 4 dimensional Gra\ss mann resp. Clifford algebra. The output is suppressed in most steps since its very tedious and long. The defining equations for the three antipodes are stored
in \verb+eq_gr+, \verb+eq_cl+ and \verb+eq_bc+.
\begin{Verbatim}
dim_V:=2:bas:=cbasis(dim_V):
B :=matrix(dim_V,dim_V,(i,j)->b[i,j]):        # scalar product
BI:=matrix(dim_V,dim_V,(i,j)->p[i,j]):        # coscalar product
make_BI_Id():                          ## <== initialize the Clifford coproduct
S   := matrix(2^dim_V,2^dim_V,(i,j)->s[i,j]): # antipode template
Sop := proc(x) linop(x,S) end:                # operator using S
X   := add(x[i]*bas[i],i=1..2^dim_V):         # general element
eq_gr :=drop_t(&map(mapop(&gco(X),1,Sop),1,wedge)-gco_unit(&t(X),1)):
eq_cl :=drop_t(&map(mapop(&gco(X),1,Sop),1,cmul )-gco_unit(&t(X),1)):
eq_bc :=drop_t(&map(mapop(&cco(X),1,Sop),1,cmul )-gco_unit(&t(X),1)):
\end{Verbatim}
In a second step we solve these equations using the $\BIGEBRA$ tangle solver \verb+tsolve1+:
\begin{Verbatim}
sol_gr :=tsolve1(clicollect(eq_gr) ,[seq(seq(s[i,j],
            i=1..2^dim_V),j=1..2^dim_V)],[seq(x[i],i=1..2^dim_V)]):
sol_cl :=tsolve1(clicollect(eq_cl) ,[seq(seq(s[i,j],
            i=1..2^dim_V),j=1..2^dim_V)],[seq(x[i],i=1..2^dim_V)]):
sol_bc :=tsolve1(clicollect(eq_bc) ,[seq(seq(s[i,j],
            i=1..2^dim_V),j=1..2^dim_V)],[seq(x[i],i=1..2^dim_V)]):
\end{Verbatim}
Before we display these morphisms in a matrix form, we compute a normalization factor for the antipode of the Clifford bi-convolution to simplify the output, which will be given as $S_{bc}= N*\,S_{bc}^\prime.$ Here the primed antipode is the actual one and the unprimed
is the normalized one.
\begin{Verbatim}
N:=linalg[det](linalg[diag](1$dim_V)- B &* BI):
S_GR := subs(sol_gr[1],evalm(S)):
S_CL := subs(sol_cl[1],evalm(S)):
S_BC := map(simplify@expand,subs(sol_bc[1],N*evalm(S))):
S_GR=evalm(S_GR),S_CL=evalm(S_CL),S_bc=map(simplify,evalm(S_BC));
\end{Verbatim}
\begin{mdisplay}
\begin{gather*}
\mathit{S\_GR}= \left[ 
\begin{array}{rrrr}
1 & 0 & 0 & 0 \\
0 & -1 & 0 & 0 \\
0 & 0 & -1 & 0 \\
0 & 0 & 0 & 1
\end{array}
\right],
\quad
\mathit{S\_CL}= \left[ 
\begin{array}{rrrc}
1 & 0 & 0 & {b_{1, \,2}} - {b_{2, \,1}} \\
0 & -1 & 0 & 0 \\
0 & 0 & -1 & 0 \\
0 & 0 & 0 & 1
\end{array}
\right],\\[2ex]
\mathit{S\_bc}= \left[ 
\begin{array}{crrc}
1 - {p_{1, \,2}}\,{b_{2, \,1}} + {p_{2, \,1}}\,{b_{2, \,1}} + {p
_{1, \,2}}\,{b_{1, \,2}} - {p_{2, \,1}}\,{b_{1, \,2}} & 0 & 0 & {
b_{1, \,2}} - {b_{2, \,1}} \\
0 & -1 & 0 & 0 \\
0 & 0 & -1 & 0 \\
{p_{1, \,2}} - {p_{2, \,1}} & 0 & 0 & 1
\end{array}
\right] 
\end{gather*}
\end{mdisplay}
\begin{proposition}
(\verb+dim_V=2+) The antipode $S_{bc}$ of a Clifford bi-convolution is up to a factor identical with the antipode $S^{\wedge}$ of the Gra\ss mann Hopf algebra if and only if the cliffordization is performed with a symmetric scalar product and a symmetric coscalar product (both derivable from a quadratic and a coquadratic form via polarization in $\text{char}\not=2).$
\end{proposition}
\begin{corollary}
The recursive formula for computing the antipode given by Milnor and Moore \citep{milnor:moore:1965a} and currently used in the Connes-Kreimer renormalization procedure cannot be applied in general for Clifford bi-convolutions with antisymmetric part in the scalar and coscalar product
\begin{gather*}
S(x) = \epsilon(x) -x - S(x^\prime_{(1)})x^\prime_{(2)},\quad S(\Id) = \Id
\end{gather*}
where the primed coproduct is over proper cuts only, i.e. $x^\prime_{(i)}\not=\Id.$
\end{corollary}
Note however that bilinear forms having antisymmetric parts are involved in the process of Wick normal ordering in QFT \citep{fauser:2001b}. A CAS is a valuable help here to find explicit examples of such structures.

\section{Deformation and quantum Yang-Baxter equation}

In this section we want to use $\BIGEBRA$ to make the relation between cliffordized products, coproducts and standard deformation theory explicit. We keep the definitions from the previous section, i.e. \verb+dim_V=2+, the settings for the scalar product \verb+B+, the coscalar product \verb+BI+, and the general element \verb+X+. For further reference we need an explicit form of the scalar product extended to $V$ and recall the matrix form of the Gra\ss mann antipode, this time from the $\BIGEBRA$ builtin function $\gantipode.$
We compute the convolutive inverse of the bilinear form needed below:
\begin{Verbatim}
BW  :=matrix(2^dim_V,2^dim_V,(i,j)->EV(cmul(bas[i],bas[j]),Id)):
S_Gr:=matrix(2^dim_V,2^dim_V,(i,j)->EV(bas[i],gantipode(bas[j]))):
BS  :=evalm(BW &* S_Gr):
BW=evalm(BW),S_gr=evalm(S_Gr),BS=evalm(BS);
\end{Verbatim}
\begin{mdisplay}
\begin{gather*}
\mathit{BW}= \left[ 
\begin{array}{rccc}
1 & 0 & 0 & 0 \\
0 & {b_{1, \,1}} & {b_{1, \,2}} & 0 \\
0 & {b_{2, \,1}} & {b_{2, \,2}} & 0 \\
0 & 0 & 0 & {b_{2, \,1}}\,{b_{1, \,2}} - {b_{2, \,2}}\,{b_{1, \,1}}
\end{array}
\right], \quad
\mathit{S\_gr}= \left[ 
\begin{array}{rrrr}
1 & 0 & 0 & 0 \\
0 & -1 & 0 & 0 \\
0 & 0 & -1 & 0 \\
0 & 0 & 0 & 1
\end{array}
\right],\\[1ex]
\mathit{BS}= \left[ 
\begin{array}{rccc}
1 & 0 & 0 & 0 \\
0 &  - {b_{1, \,1}} &  - {b_{1, \,2}} & 0 \\
0 &  - {b_{2, \,1}} &  - {b_{2, \,2}} & 0 \\
0 & 0 & 0 & {b_{2, \,1}}\,{b_{1, \,2}} - {b_{2, \,2}}\,{b_{1, \,1}}
\end{array}
 \right]
\end{gather*}
\end{mdisplay}
Note that the convolutive inverse is just the scalar product w.r.t the negative bilinear form.
\begin{proposition}
Not every endomorphism has as the convolutive inverse $B^S= B\circ S.$ However, exponentially generated morphisms and, more generally, graded morphisms do.
\end{proposition}

A quasi triangular structure is an element $R \in V\otimes V$ which satisfies, among others, the following condition: 
\[
\hat{\text{sw}}\circ\Delta_{Cl}(x) = (R_{(1)}\otimes R_{(2)})\circ \Delta_{Gr}(x)
\tag{$*$}
\]
see, e.g., \citet{majid:1995a}. Note that our definition is somewhat different due to our reversed ordering of tensor products of dual elements. The task is now to define a general element $R$ and try to compute all possible solutions of equation $(*).$
Therefore we start to define $R$ as
\begin{Verbatim}
Y:=add(y[i]*bas[i],i=1..2^dim_V):
R:=proc(x,y) local bas,tr_tbl; option remember;
   bas:=cbasis(dim_V):
   tr_tbl:=table([seq(op(Cliff5[extract](bas[i]))=i,i=1..2^dim_V)]);
   R[tr_tbl[op(Cliff5[extract](x))],tr_tbl[op(Cliff5[extract](y))]]
end:
RR:=add(add(contract(&t(bas[i],bas[j]),1,R)*&t(bas[i],bas[j]),
   i=1..2^dim_V),j=1..2^dim_V):
\end{Verbatim}
Now the actual computation starts by evaluating the l.h.s. and r.h.s. of $(*)$ as
\begin{Verbatim}
Leq:=gswitch(&cco(X),1):
Req:=&map(&map(switch(switch(&t(RR,&gco(X)),2),1),3,wedge),1,wedge):
\end{Verbatim}
Note that in the right hand side of the equation \verb+Req+, two switches were used to put $R$ in-between the $x_{(i)}$ of the coproduct and are not part of the formula $(*)$ while the graded switch in the l.h.s. equation \verb+Leq+ is generic. This time we show directly how to solve tangle equations:
\begin{Verbatim}
eq:=tcollect(Leq-Req):
vars:={seq(seq(R[i,j],i=1..2^dim_V),j=1..2^dim_V)}:
T:={seq(seq(&t(bas[i],bas[j]),i=1..2^dim_V),j=1..2^dim_V)}:
CO:={seq(x[i],i=1..2^dim_V)}:
sys:={}:
for t in T do  
for co in CO do
  sys:={op(sys),coeff(coeff(eq,t),co)};
od:od:
sol:=solve(sys,vars):
matR:=matrix(2^dim_V,2^dim_V,(i,j)->R[i,j]):
`R`=subs(sol,evalm(matR));  
\end{Verbatim}
\begin{mdisplay}
\[
R= \left[ 
{\begin{array}{rccc}
1 & 0 & 0 & 0 \\
0 &  - {p_{1, \,1}} &  - {p_{2, \,1}} & 0 \\
0 &  - {p_{1, \,2}} &  - {p_{2, \,2}} & 0 \\
0 & 0 & 0 & {p_{2, \,1}}\,{p_{1, \,2}} - {p_{2, \,2}}\,{p_{1, \,1
}}
\end{array}}
 \right] 
\]
\end{mdisplay}
We have put the solution once more in matrix form. Inspection of this result shows, that our $R$ is the convolutive inverse of the coscalar product, and hence it is exponentially generated. In fact it is an axiom of a quasi triangular structure that a convolutive inverse exists. We can check our assertion explicitly via
\begin{Verbatim}
eq1:=gco_unit(gco_unit(&t(X,Y),2),1):
eq2:=simplify(drop_t(contract(contract(&t(&gco(X),&gco(Y))
                            ,2,scalarpart@cmul[-K]),1,scalarpart@cmul[K])))*Id:
is(eq1=eq2);
\end{Verbatim}
\begin{mdisplay}
\[
\mathit{true}
\]
\end{mdisplay}
\begin{corollary}
The (co-)cliffordizations with respect to (co-)bilinear forms $-K$ and $K$ are convolutive inverse w.r.t. the Gra\ss mann Hopf convolution.
\end{corollary}

We are now ready to check that our quasi triangular structure fulfills the {\em quantum Yang Baxter equation}. We do this first for the graded switch of the Gra\ss mann Hopf convolution, which is trivially a braid. We compute
\begin{Verbatim}
Z:=add(z[i]*bas[i],i=1..2^dim_V): eq0:=&t(X,Y,Z):
eq1:=gswitch(gswitch(gswitch(eq0,1),2),1):
eq2:=gswitch(gswitch(gswitch(eq0,2),1),2):
is(eq1=eq2);
\end{Verbatim}
\begin{mdisplay}
\[
\mathit{true}
\]
\end{mdisplay}
where we have not shown the cumbersome terms but simply checked that both sides evaluate to the same result, which establishes the assertion. 

Since a quasi triangular structure is an exponentially generated endomorphism, we have to check that it fulfills
$$
R_{12}R_{23}R_{12}=R_{23}R_{12}R_{23}.
$$
To check this assertion we calculate with $\BIGEBRA:$
\begin{Verbatim}
bw:=proc(x,y) local i,j;
 add(add(BW[i,j]*EV(bas[i],y)*EV(bas[j],x),i=1..2^dim_V),j=1..2^dim_V) 
end:
Bsw:=proc(x,i)
  tcollect(gswitch(contract(&gco(&gco(x,i+1),i),i+1,bw),i))
end:
eq3:=Bsw(Bsw(Bsw(eq0,1),2),1):
eq4:=Bsw(Bsw(Bsw(eq0,2),1),2):
is(eq3=eq4);
\end{Verbatim}
\begin{mdisplay}
\[
\mathit{true}
\]
\end{mdisplay}
which proves our claim in \verb+dim_V=2+. As a last demonstration, we check the remaining properties of a quasi triangular structure to be valid for an exponentially generated endomap.
\begin{gather*}
R(S(a),b) = R^S(a,b), \quad R^{S}(a,S(b)) = R(a,b),\quad R(S(a),S(b))=R(a,b).
\end{gather*}
Hence we generate the endomap `R' and apply to its arguments the Gra\ss mann antipode:
\begin{Verbatim}
R:='R':
out[1]:=`R`=matrix(2^dim_V,2^dim_V,(i,j)->scalarpart(cmul[R](bas[i],bas[j]))):
out[2]:=`RS`=matrix(2^dim_V,2^dim_V,(i,j)->scalarpart(cmul[R]
                                                 (gantipode(bas[i]),bas[j]))):
out[1];out[2];
\end{Verbatim}
\begin{mdisplay}
\begin{gather*}
R= \left[ 
\begin{array}{rccc}
1 & 0 & 0 & 0 \\
0 & {R_{1, \,1}} & {R_{1, \,2}} & 0 \\
0 & {R_{2, \,1}} & {R_{2, \,2}} & 0 \\
0 & 0 & 0 & {R_{2, \,1}}\,{R_{1, \,2}} - {R_{2, \,2}}\,{R_{1, \,1
}}
\end{array}
 \right]\\[1ex]
\mathit{RS}= \left[ 
\begin{array}{rccc}
1 & 0 & 0 & 0 \\
0 &  - {R_{1, \,1}} &  - {R_{1, \,2}} & 0 \\
0 &  - {R_{2, \,1}} &  - {R_{2, \,2}} & 0 \\
0 & 0 & 0 & {R_{2, \,1}}\,{R_{1, \,2}} - {R_{2, \,2}}\,{R_{1, \,1
}}
\end{array}
 \right] 
\end{gather*}
\end{mdisplay}
\begin{Verbatim}
out[3]:=`RSS`=matrix(2^dim_V,2^dim_V,(i,j)->scalarpart(cmul[R]
                                      (gantipode(bas[i]),gantipode(bas[j])))):
out[4]:=`SR`=matrix(2^dim_V,2^dim_V,(i,j)->scalarpart(cmul[R]
                                                 (bas[i],gantipode(bas[j])))):
out[3];out[4];
\end{Verbatim}
\begin{mdisplay}
\begin{gather*}
\mathit{RSS}= \left[ 
\begin{array}{rccc}
1 & 0 & 0 & 0 \\
0 & {R_{1, \,1}} & {R_{1, \,2}} & 0 \\
0 & {R_{2, \,1}} & {R_{2, \,2}} & 0 \\
0 & 0 & 0 & {R_{2, \,1}}\,{R_{1, \,2}} - {R_{2, \,2}}\,{R_{1, \,1}}
\end{array}
\right]\\[1ex] 
\mathit{SR}= \left[ 
\begin{array}{rccc}
1 & 0 & 0 & 0 \\
0 &  - {R_{1, \,1}} &  - {R_{1, \,2}} & 0 \\
0 &  - {R_{2, \,1}} &  - {R_{2, \,2}} & 0 \\
0 & 0 & 0 & {R_{2, \,1}}\,{R_{1, \,2}} - {R_{2, \,2}}\,{R_{1, \,1}}
\end{array}
\right] 
\end{gather*}
\end{mdisplay}
Comparing the matrix representations gives us the desired results when we recall that the convolutive inverse is given by the `scalar product` \verb+-R+ on the generating space.

We can compute along the same lines as exemplified above the {Yang Baxter matrix}, a 16 by 16 matrix, which represents the action of $R$ on $V\otimes V.$ Since the output of these computations is quite lengthy we will give only some further results and not the explicit matrix. The actual worksheet is available from the home page of the second author.

\begin{remark}
The Yang Baxter matrix of our 2 dimensional example has determinant 1 and is a function of all four parameters of the quasi triangular structure $R.$
\end{remark}

A further question is, if this structure is really quasi triangular or only triangular. Due to our reversed indexing in duals, we have to check for triangularity if the Yang Baxter matrix squares to unity. 

\begin{remark}
The Yang Baxter matrix of our 2 dimensional example is triangular if and only if the exponentially generated endomap is derived from a symmetric bilinear form on the generating space. This has deep implications for ordering process in quantum field theory, see \citet{fauser:2001b}.
\end{remark}
 
We expect such assertions to be true in higher dimensions, though that needs an algebraic proof. However, having solutions in low dimensions allows in many cases to prove by induction a general hypothesis. This is the step from experimental mathematics to `pure' mathematics.

\section{Conclusions}

The present paper was intended to show how a CAS allows to ask questions of research interest and to come up with low dimensional solutions. During this process we have worked out examples showing how to use the $\BIGEBRA$ package and how to attack more advanced problems. 

We hope that it became obvious from our presentation of the $\CLIFFORD$ \citep{ablamowicz:fauser:2002c} and $\BIGEBRA$ packages that the ability to compute, via a CAS, allows to build a sound knowledge about mathematical problems and that there is a need for {\em experimental mathematics} in education and research. 

This article is, due to restrictions in space, obviously not able to give a complete review of all abilities of the $\CLIFFORD$ and $\BIGEBRA$ packages. Therefore the interested reader is invited to check the package home page\footnote{url: {\small\tt http://math.tntech.edu/rafal/}} for the documentation. Both packages come with a built-in online help where, for every function, syntax, synopsis and examples, and sometimes a more advanced mathematical background is provided. A printed ps or pdf version of approx. 500 pages is available there also. For those who wish to have a look at and a feel for the packages, there is a possibility to access them online over the web\footnote{url: {\small\tt http://kaluza.physik.uni-konstanz.de/\~\ fauser}}. 

There are tremendously many open problems coming with a bi-convolution of mathematical and physical type. It is not yet clear, under which
conditions antipodes exist in higher dimensions. In dimension 3 a Clifford bi-convolution has in general no antipode, hence further relations between scalar and coscalar product have to be found. Only very little is known about the nature of the crossings derived from antipodal bi-convolution, but see \citet{fauser:oziewicz:2001a}. The axiomatics, in terms of morphims and category theory of Clifford bi-convolutions, is under consideration but not yet finished \citep{oziewicz:1998a,oziewicz:2001a}. Clifford bi-convolutions are deeply connected to the renormalization process of the quantum filed theory. However, there one has to go beyond the scheme presented in this paper, see \citet{brouder:2002a,brouder:fauser:frabetti:oeckl:2002a}. A CAS and especially $\CLIFFORD$ and $\BIGEBRA$ are ideally suited to explore this exciting field.

Finally we want to emphasize that the present article contains Maple output (with only a minor \TeX\ cosmetics) which was generated directly by processing a Maple worksheet.   

\vskip+15pt
\noindent{\bf Acknowledgment:}
The second author, BF, acknowledges gratefully financial support from University of Konstanz, LS Prof. Heinz Dehnen.

\vskip 2em
\noindent Submitted: \today; Revised: TBA.
\end{document}